# A Variant of the Point Defect Model for Passivity of Metals

Bosco Emmanuel, Modeling and Simulation, Central Electrochemical Research Institute, Karaikudi-630006, India

**Abstract**

A variant of the point defect model originally enunciated by Macdonald and co-workers is advanced and its theoretical implications for the steady state current density, barrier layer thickness and the concentration of metal vacancy at the metal/film interface are deduced. The differences between the original point defect model and the present variant are also highlighted. The empirical parameters $\alpha$ and $\beta$ in the original point defect model are replaced with two physical parameters $R_{cont}$ and $\rho_f$ which represent respectively the electronic contact resistance at the metal/film interface and the electronic resistivity of the oxide film. The present variant correctly describes the annihilation of the metal vacancies at the metal/film interface and also enforces the conservation of particle and defect volumes during the solid-state reactions leading to the natural inclusion of the famous Pilling-Bedsworth ratio $R_{PB}$ into the model. Diagnostics which help to check the model predictions with experiments are given. Use of this variant to describe strees-induced failure of the barrier oxide leading to pitting is also discussed.

**Introduction**

The point defect model was first proposed in 1981 and has progressed considerably over the past 30 years both in its theoretical framework and in its applications to analyse experimental data [1]. It holds much promise for the eventual complete understanding the passive state of a metal $M$ with an oxide layer $MO_{\chi/2}$ on it. It also surpassed in several ways many earlier attempts by physicists. Macdonald and co-workers have made useful comparisons between the point defect model and these earlier models. The objective of the present work is to present a variant of the point defect model, deduce its predictions for the passive current density, the barrier layer thickness and the metal vacancy concentrations all at steady state and provide the experimental diagnostics based on the applied potential and the pH. Unlike the original point defect model the proposed variant does not contain any empirical parameters like the $\alpha$ and $\beta$ in the original model. All model parameters have a physical basis. Besides, one of the 7 defect reactions postulated by Macdonald et al is modified into a more appropriate form and we show that this modification has non-trivial implications for the model predictions. Before we proceed to the variant, the important assumptions of the original point defect model are stated below:

(i) The barrier oxide layer is rich in cation vacancies & interstitials and anion vacancies with the defect concentrations greater than in the isolated bulk oxide.

(ii) The defects are continuously generated and/or annihilated at the metal/film and at the film/solution interfaces.
(iii) The electric field strength $\varepsilon$ in the oxide film is independent of the applied voltage and remains constant throughout the oxide film. This voltage independence is argued to arise from band-to-band Esaki tunneling.
(iv) The electrical potential drops across the metal/film/solution sandwich are assumed to be of the from:

$$\phi_{f/s} = \alpha.V + \beta.pH + \phi^0_{f/s} \tag{1}$$

$$\varepsilon.L = the.potential.drop.across.the.oxide.film.of.thickness.L \tag{2}$$

and $$\phi_{m/f} = (1-\alpha).V - \beta.pH - \phi^0_{f/s} - \varepsilon.L + \phi_R \tag{3}$$

where $\alpha$ and $\beta$ are two empirical parameters to be evaluated by fitting the experimental data to the model and all other parameters are as defined in the original papers by Macdonald and co-workers.

The defect reactions postulated by Macdonald et al are:

    Metal  |  Barrier Layer  |  Porous Layer OR Solution

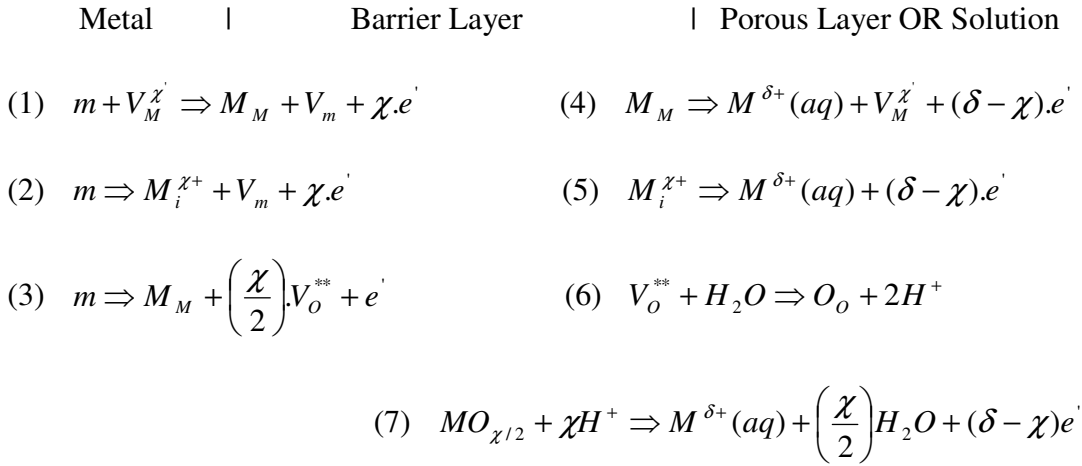

(1) $m + V^{\chi'}_M \Rightarrow M_M + V_m + \chi.e'$      (4) $M_M \Rightarrow M^{\delta+}(aq) + V^{\chi'}_M + (\delta - \chi).e'$

(2) $m \Rightarrow M^{\chi+}_i + V_m + \chi.e'$      (5) $M^{\chi+}_i \Rightarrow M^{\delta+}(aq) + (\delta - \chi).e'$

(3) $m \Rightarrow M_M + \left(\dfrac{\chi}{2}\right).V^{**}_O + e'$      (6) $V^{**}_O + H_2O \Rightarrow O_O + 2H^+$

(7) $MO_{\chi/2} + \chi H^+ \Rightarrow M^{\delta+}(aq) + \left(\dfrac{\chi}{2}\right)H_2O + (\delta - \chi)e'$

The rate constants for these 7 reactions are respectively $k_1$ to $k_7$.

**The Variant**

The variant of the PDM studied in the present work retains the two basic assumptions (i) and (ii), drops assumption (iii), and assumption (iv) is replaced by

$$\phi_{m/f} = R_{cont}.i + \phi^{oc}_{m/f} \tag{4}$$

$$\varepsilon.L = \rho_f.i.L \tag{5}$$

$$\phi_{f/s} = V_{ap} - R_{cont}.i - \phi^{oc}_{m/f} - \rho_f.i.L - \phi_R \tag{6}$$

Here $i$ is the current density through the system, $R_{cont}$ is the ohmic contact resistance at the metal/film interface which can also be interpreted as arising from the linearised Schottky diode equation $i = i_{sat}.(\exp[F.(\phi_{m/f} - \phi^0_{m/f})/RT] - 1)$ where $i_{sat}$ is the saturation current density of the diode is and $\phi^0_{m/f}$ is the zero-current diode voltage known as the built-in potential in Semiconductor Physics. In this interpretation $R_{cont} = \dfrac{RT}{F.i_{sat}}$. $\rho_f$ is the electronic resistivity of the oxide film. It is to be noted that even an uncompensated ohmic resistance in the outer porous layer or in the solution may easily be included in the equation (6) for $\phi_{f/s}$ without ever changing the model.

The defect reactions considered in the present variant are:

Metal  |  Barrier Layer  |  Porous Layer OR Solution

(1) $m + V_M^{\chi'} \Rightarrow M_M + V_m + \chi.e'$

(4) $M_M \Rightarrow M^{\delta+}(aq) + V_M^{\chi'} + (\delta - \chi).e'$

(2) $m + V_{M_i} \Rightarrow M_i^{\chi+} + V_m + \chi.e'$

(5) $M_i^{\chi+} \Rightarrow M^{\delta+}(aq) + V_{M_i} + (\delta - \chi).e'$

(3') $m + qV_m + \left(\dfrac{\chi}{2}\right)O_O \Rightarrow MO_{\chi/2} + \left(\dfrac{\chi}{2}\right)V_O^{**} + e'$

(6) $V_O^{**} + H_2O \Rightarrow O_O + 2H^+$

(7) $MO_{\chi/2} + \chi H^+ \Rightarrow M^{\delta+}(aq) + \left(\dfrac{\chi}{2}\right)H_2O + (\delta - \chi)e'$

This set of reactions is the same as the set proposed by Macdonald et al and given above in the Introduction except reaction (3) where the metal vacancy $V_m$, $O_O$ and $MO_{\chi/2}$ are the additional species. Figure 1 provides a pictorial comparison of reaction (3) of Macdonald et al and reaction (3') of the present variant. Refering to the pictorial representation of reaction (3)[of the original point defect model] in Figure 1, it is unclear how $\dfrac{\chi}{2}.V_O^{**}$ is generated. This $\dfrac{\chi}{2}.V_O^{**}$ seems to be the pre-existing anion vacancies in the oxide layer and hence can NOT be said to be generated by reaction 3. This is to be compared with the pictorial for reaction 3'[of the present variant] where it is clear that $\dfrac{\chi}{2}.V_O^{**}$ is produced by $\dfrac{\chi}{2}.O_O$ which leaves the oxide phase and lodges in the metal vacancies $V_m$ in the metal phase.

**Figure 1 <u>Pictorial Representations of Reactions 3 and 3'</u>**

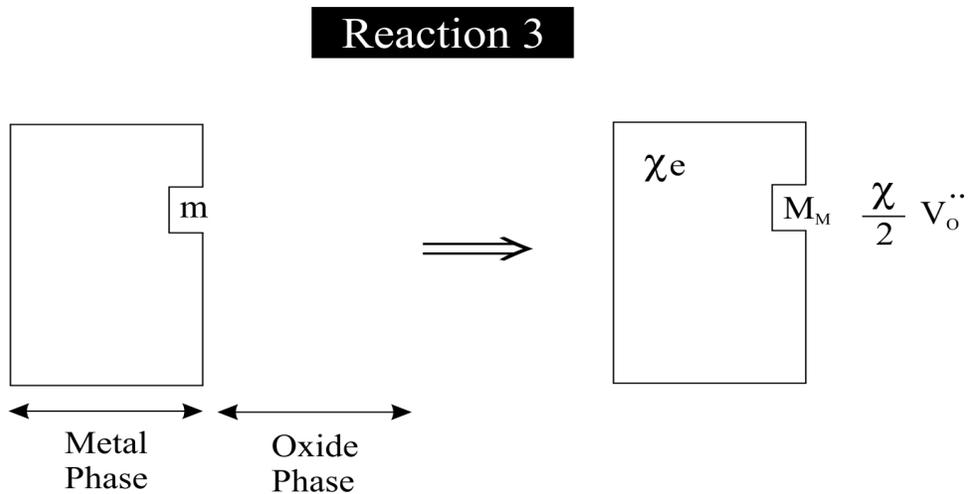

**Reaction 3 in the original point defect model**

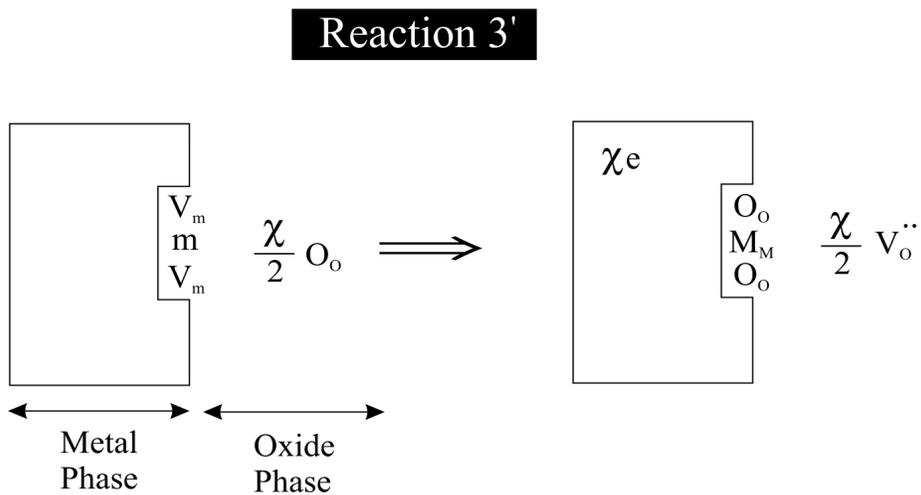

**Reaction 3' in the present variant of the point defect model**

There is also a difficulty in establishing the steady state if we do not invoke $V_m$ and $O_O$ [as in reaction 3'] because the rate of generation of $V_m$ in reactions 1 and 2 will not be balanced by its consumption and the rate of generation of $O_O$ at film/solution interface in reaction 6 will not be balanced by its consumption at metal/film interface. Reactions 2 and 5 of the variant are essentially the corresponding older reactions except that the vacant cation interstitial $V_{M_i}$ is explicitly shown. While this change does not have any consequences for the model predictions, the change contained in reaction 3' has important consequences for the model predictions. In the terminology used by Macdonald et al reaction 1, 2, 4, 5 and 6 are lattice-conserving while reactions 3' and 7 are lattice non-conserving. Another aspect to be considered while writing defect reactions in the solid is volume conservation: the sum of the effective volumes of particles and vacancies before the reaction must equal the sum of the effective volumes of particles and vacancies produced by the reaction. Thus one more conservation principle is added to the usual mass balance and charge balance of reactions. Applying this volume conservation principle to reaction 3', we obtain

$$\text{Volume of } m + q \cdot (\text{Volume of } V_m) + \left(\frac{\chi}{2}\right) \cdot (\text{Volume of } O_O) = (\text{Volume of } MO_{\chi/2})$$
$$+ \left(\frac{\chi}{2}\right)(\text{Volume of } V_O^{**})$$

….. (7)

If we approximately set the volume of a particle equal to the corresponding defect volume, it turns out that

$$q = \frac{Volume.of.MO_{\chi/2}}{Volume.of.m} - 1 \quad (8)$$

$$= R_{PB} - 1 \quad (9)$$

where $R_{PB}$ is the famous Pilling-Bedworth ratio. It is satisfying to note that this number which was proposed in the olden times as an index of anodic film stability enters through reaction 3' of the present variant of the point defect model.

The reaction rate equations for the defect reactions 1 to 7 may now be written. The concentrations of the majority species $M_M$, $O_O$, $m$, $V_{M_i}$ and $MO_{\chi/2}$ will be absorbed into the rate constants themselves while the concentrations of the minority species $V_M^{\chi'}, V_O^{**}, M_i^{\chi+}$ and $V_m$ are shown explicitly in the rate expressions:

$$R_1 = k_1.C_M(0) \qquad\qquad R_4 = k_4 \qquad\qquad (10)$$

$$R_2 = k_2 \qquad\qquad R_5 = k_5.C_{M_i}(L) \qquad\qquad (11)$$

$$R_3 = k_3.\left(\frac{C_m(0)}{C_m^0}\right)^k \qquad\qquad R_6 = k_6.C_O(L) \qquad\qquad (12)$$

$$R_7 = k_7.\left(\frac{C_{H^+}}{C_{H^+}^0}\right)^n \qquad\qquad (13)$$

where $C_M(0)$ and $C_m(0)$ are respectively the concentrations of $V_M^{\chi'}$ and $V_m$ at $x=0$ (i.e. at the metal/film interface), $C_{M_i}(L)$ and $C_O(L)$ are the concentrations of $M_i^{\chi+}$ and $V_O^{**}$ at $x=L$ (i.e. at the film/solution interface). The unit of $k_1$, $k_5$ and $k_6$ is $cm.\sec^{-1}$ while that of $k_2$, $k_3$, $k_4$ and $k_7$ is $mole.cm^{-2}.\sec^{-1}$.

The rate constants $k_1$ through $k_7$ are related to the interfacial potential drops $\phi_{m/f}$ and $\phi_{f/s}$ at $x=0$ and $x=L$ as follows:

$$k_1 = k_1^0.\exp[\alpha_1.\chi.F.(\phi_{m/f} - \phi_{m/f}^{oc})/RT] \qquad\qquad (14)$$

$$k_2 = k_2^0.\exp[\alpha_2.\chi.F.(\phi_{m/f} - \phi_{m/f}^{oc})/RT] \qquad\qquad (15)$$

$$k_3 = k_3^0.\exp[\alpha_3.\chi.F.(\phi_{m/f} - \phi_{m/f}^{oc})/RT] \qquad\qquad (16)$$

$$k_4 = k_4^0.\exp[\alpha_4.\delta.F.(\phi_{f/s} - \phi_{f/s}^{oc})/RT] \qquad\qquad (17)$$

$$k_5 = k_5^0.\exp[\alpha_5.\delta.F.(\phi_{f/s} - \phi_{f/s}^{oc})/RT] \qquad\qquad (18)$$

$$k_6 = k_6^0.\exp[2\alpha_6.F.(\phi_{f/s} - \phi_{f/s}^{oc})/RT] \qquad\qquad (19)$$

$$k_7 = k_7^0.\exp[\alpha_7.(\delta-\chi).F.(\phi_{f/s} - \phi_{f/s}^{oc})/RT] \qquad\qquad (20)$$

where $\phi_{m/f}$ and $\phi_{f/s}$ are given by equations (4) and (6) and the $k_i^0$'s are the standard rate constants related to but not the same as the base rate constants defined by Macdonald and co-workers.

**Expressions for the steady state current $i_{ss}$ and barrier layer thickness $L_{ss}$ and the concentration of metal vacancy $V_m$**

The steady state current density $i_{ss}$ is given by

$$i_{ss} = F.\{-\delta.J_M + \delta.J_i + 2.J_O + (\delta - \chi).R_7\} \tag{21}$$

where $J_M$, $J_i$ and $J_O$ are respectively the flux of the species $V_M^{\chi'}$, $M_i^{\chi+}$ and $V_O^{**}$. $R_7$ is given by equation (13). Now the rate of destruction of the oxide layer by reaction 7 equals at steady state the rate of formation of the oxide layer by reaction 3'. Therefore

$$R_7 = \left(\frac{2}{\chi}\right).J_O \tag{22}$$

$$\therefore\ i_{ss} = \delta.F.\{-J_M + J_i + R_7\} \tag{23}$$

Further, for sustaining the steady state, the rate of production of the metal vacancy $V_m$ by reactions 1 and 2 should equal the rate of its annihilation by reaction 3. Therefore

$$J_i - J_M = \left(\frac{q}{(\chi/2)}\right).J_O \tag{24}$$

Using equation (22) this becomes

$$J_i - J_M = q.R_7 \tag{25}$$

$$\therefore\quad i_{ss} = \delta.F.R_{PB}.R_7 \tag{26}$$

It is interesting to note that the present variant predicts a simple equation for the steady state current which depends only on the rate of dissolution of the oxide layer $R_7$. This form differs from the form predicted by the original point defect model. The difference is directly attributable to the use of reaction 3' in the place of reaction 3.

At this point we need to distinguish between two cases A and B. Case A will use the equations (1) to (3) for the potential drops which involve the empirical parameters $\alpha$ and $\beta$ of Macdonald et al and Case B will use equations (4) to (6) proposed in the present work which involve the physical parameters $R_{cont}$ and $\rho_f$. Nonetheless the use of reaction 3' in the place of reaction 3 is common for Case A and Case B. Case A and Case B has two further sub-cases: $\delta = \chi$ and $\delta \neq \chi$.

**Case A**
(i) For $\delta = \chi$ the steady state current is simply

$$i_{ss} = \delta.F.R_{PB}.k_7^0.C_R^n \qquad (27)$$

Where and all through this paper

$$C_R = \frac{C_{H^+}}{C_{H^+}^0}$$

Clearly the steady state current density is independent of potential and depends only on pH.

(ii) For $\delta \neq \chi$ the steady state current is given by

$$i_{ss} = \delta.F.R_{PB}.k_7^0.\exp\{a_7.V + c_7.pH\}C_R^n \qquad (28)$$

Which depends on both potential and pH.

Steady state barrier layer thickness $\delta \neq \chi$

Use the equation

$$J_i - J_M = q.R_7 \qquad (29)$$

to obtain the following analytic form for $L_{ss}$

$$L_{ss} = \left(\frac{1}{b_2}\right).\ln\left(\frac{k_2^0 \exp\{a_2.V + c_2.pH\}}{q.k_7^0.C_R^n.\exp\{a_7.V + c_7.pH\} - k_4^0.\exp\{a_4.V + c_4.pH\}}\right) \qquad (30)$$

Where the $a_i$'s and the $b_i$'s are as defined by Macdonald et al.

For the special case when only cation interstitials and anion vacancies dominate, this reduces to:

$$L_{ss} = \left(\frac{a_2 - a_7}{b_2}\right).V + \left(\frac{c_2 - c_7}{b_2}\right).pH + \left(\frac{1}{b_2}\right).\ln\left(\frac{k_2^0}{q.k_7^0.C_R^n}\right) \qquad (30')$$

which is linear in the potential.

**Case B**

(i) For $\delta = \chi$ the steady state current is the same as in Case A

(ii) For $\delta \neq \chi$, it is convenient to start with the two equations:

$$i_{ss} = \delta.F.R_{PB}.R_7 \tag{31}$$

and

$$J_i - J_M = q.R_7 \tag{32}$$

to obtain

$$J_i - J_M = q.\frac{i_{ss}}{\delta.F.R_{PB}} \tag{33}$$

Now, using the explicit forms for $J_i$, $J_M$ and $i_{ss}$, we obtain after some algebraic steps:

$$i_{ss} = \left(\frac{\delta.F.R_{PB}}{q}\right).\{k_2^0.\exp(\alpha_2.\chi.F.R_{cont}.i_{ss}/RT) + k_4^0.\left(\frac{i_{ss}}{\delta.F.R_{PB}.k_7^0.C_R^n}\right)^P\} \tag{34}$$

where $P = \dfrac{\alpha_4.\delta}{\alpha_7(\delta - \chi)}$ \hfill (35)

Though this is a non-linear equation to be solved for $i_{ss}$, it is interesting to note that it is independent of the potential and depends only on the pH.

**Steady state barrier layer thickness $L_{ss}$**

For $\delta = \chi$ we again start from equation (33) and obtain the following analytical formula for $L_{ss}$:

$$L_{ss} = \frac{V_{ap} - V_{oc}}{\rho_f.i_{ss}} - \frac{R_{cont}}{\rho_f} - \left(\frac{RT}{\alpha_4.\delta.F.\rho_f.i_{ss}}\right).\ln\{\left(\frac{q.i_{ss}}{k_4^0.\delta.F.R_{PB}}\right) - \left(\frac{k_2^0}{k_4^0}\right).\exp(\alpha_2.\chi.F.R_{cont}.i_{ss}/RT)\}$$

….. (36)

Note that $L_{ss}$ is linear in $V_{ap}$ and depends on pH only through $i_{ss}$.

For $\delta \neq \chi$ we once again start from equation (33). However we obtain a different form of $L_{ss}$ as the form of $i_{ss}$ is different now.

$$L_{ss} = \frac{V_{ap} - V_{oc}}{\rho_f . i_{ss}} - \frac{R_{cont}}{\rho_f} - \left( \frac{RT}{\alpha_7 .(\delta - \chi).F.\rho_f .i_{ss}} \right) . \ln\left( \frac{i_{ss}}{\delta.F.R_{PB}.k_7^0.C_R^n} \right) \tag{36}$$

Again $L_{ss}$ is linear in $V_{ap}$, though now depends on pH directly besides through $i_{ss}$.

**Steady metal vacancy concentration at the metal/film interface**

The steady state assumption can also be used to compute the steady state metal vacancy concentration $C_m(0)$ at the metal/film interface. The steady state requires that $J_O$, and of course all other fluxes, remains constant in time. As $J_O$ depends on $C_m(0)$ as

$$J_O = \left(\frac{\chi}{2}\right).k_3.\left(\frac{C_m(0)}{C_m^0}\right)^k \tag{37}$$

the steady state requires that $C_m(0)$ remains constant in time. The only way to achieve this is to impose that:

Rate of creation of $V_m$ = Rate of annihilation of $V_m$ \hfill (38)

$$\therefore \quad J_i - J_M = q.k_3.\left(\frac{C_m(0)}{C_m^0}\right)^k \tag{39}$$

$$\therefore \quad \frac{C_m(0)}{C_m^0} = \left(\frac{J_i - J_M}{q.k_3}\right)^{1/k} \tag{40}$$

Note that $J_i = k_2$ and $J_M = -k_4$. $k_2$, $k_3$ and $k_4$ are known functions of $V$, $pH$ and $L_{ss}$ for Case A and they are known functions of $V$, $i_{ss}$ and $L_{ss}$ for Case B. Hence we insert in equation (40) the formulae for $i_{ss}$ and $L_{ss}$ which was found in the earlier sections and thereby compute the concentration of metal vacancy at the steady state. $C_m^0$ may conveniently be taken as the equilibrium concentration of metal vacancy in the metal at STP. The super-saturation $S$ of the metal vacancy then becomes

$$S = C_m(0)/C_m^0 \tag{41}$$

Note that the super-saturation controls the nucleation and growth of metal vacancies into micro-voids which in turn may lead to build-up of mechanical stresses at the metal/film interface leading to failure of the oxide barrier film and pitting. Hence the variant of the point defect model developed in the present work can provide the theoretical framework for studying stress-induced failure of the barrier oxide layer and pitting.

**Diagnostics**

From the results of the previous sections simple diagnostics follow for the way $Log(i_{ss})$ and $L_{ss}$ should vary with the potential and the pH for Case A and Case B and for the sub-cases $\delta = \chi$ and $\delta \neq \chi$.

**Case A**

$$\frac{\partial \log(i_{ss})}{\partial pH} = -n \quad \text{and} \quad \frac{\partial \log(i_{ss})}{\partial V} = 0 \quad \text{for} \quad \delta = \chi \tag{42}$$

$$\frac{\partial \log(i_{ss})}{\partial pH} = \left(\frac{c_7}{2.303} - n\right) \quad \text{and} \quad \frac{\partial \log(i_{ss})}{\partial V} = \frac{a_7}{2.303} \quad \text{for} \quad \delta \neq \chi \tag{43}$$

$$\frac{\partial L_{ss}}{\partial pH} = \left(\frac{c_2 - c_7 + 2.303 \cdot n}{b_2}\right) \quad \text{and} \quad \frac{\partial L_{ss}}{\partial V} = \left(\frac{a_2 - a_7}{b_2}\right) \tag{44}$$

This diagnostic for $L_{ss}$ is for the special case when the cation interstitials and anion vacancies dominate and holds for $\delta = \chi$ and $\delta \neq \chi$.

**Case B**

**(i)** $\delta = \chi$

$$\frac{\partial \log(i_{ss})}{\partial pH} = -n \quad \text{and} \quad \frac{\partial \log(i_{ss})}{\partial V_{ap}} = 0 \tag{45}$$

$$\frac{\partial L_{ss}}{\partial pH} \Rightarrow non-trivial \quad \text{and} \quad \frac{\partial L_{ss}}{\partial V_{ap}} = \frac{1}{\rho_f . \delta . F . R_{PB} . k_7^0 . C_R^n} \qquad (46)$$

**(ii)** $\delta \neq \chi$

$$\frac{\partial \log(i_{ss})}{\partial pH} \Rightarrow non-trivial \quad \text{and} \quad \frac{\partial \log(i_{ss})}{\partial V_{ap}} = 0 \qquad (47)$$

$$\frac{\partial L_{ss}}{\partial pH} \Rightarrow non-trivial \quad \text{and} \quad \frac{\partial L_{ss}}{\partial V_{ap}} = \frac{1}{\rho_f . i_{ss}} \qquad (48)$$

where in equation (48) $i_{ss}$ is a non-trivial function of pH.

**Conclusion**

The electrochemical impedance response of the variant studied in the present work is planned for our future work as it will provide estimates of the model parameters used. Comparisons of the model predictions with the available experimental data should throw more light on the present variant vis-à-vis the original point defect model.

**References**

**Macdonald D.D., Electrochimica Acta, vol.56, 2011, 1761-1772 [see also the long list of useful references given therein]**